\documentclass[10pt,twocolumn]{article}
\usepackage{times}
\usepackage{fullpage}
\usepackage{amsmath,amssymb,epsfig}
\usepackage{url} 
\usepackage{multirow}
\usepackage{subfigure}
\usepackage{graphicx}
\usepackage{slashbox}
\usepackage{listings}
\usepackage[linesnumbered,ruled,vlined]{algorithm2e}
\usepackage[nolineno,noindent,norules]{lgrind}

\makeatletter
\def\v#1{{\fontfamily{cmtt}\fontsize{\f@size}{\f@size}\selectfont #1}}
\newcommand{\eg}{e.g.}

\newcommand{\xxx}{\mbox{\textsc{eos}}\xspace}

\begin{document}

\title{\bf \xxx: Automatic, In-vivo Evolution of Kernel Policies
  \\ for Better Performance}
\author{Yan Cui, Quan Chen, Junfeng Yang \\ Columbia University}
\date{}
\maketitle
\thispagestyle{empty}

\begin{abstract}

Today's monolithic kernels often implement a small, fixed set of policies
such as disk I/O scheduling policies, while exposing many parameters to
let users select a policy or adjust the specific setting of the policy.
Ideally, the parameters exposed should be flexible enough for users to
tune for good performance, but in practice, users lack domain knowledge of
the parameters and are often stuck with bad, default parameter settings.

We present \xxx, a system that bridges the knowledge gap between kernel
developers and users by automatically evolving the policies and parameters
\emph{in vivo} on users' real, production workloads. It provides a simple
\emph{policy specification API} for kernel developers to programmatically
describe how the policies and parameters should be tuned, a \emph{policy
  cache} to make in-vivo tuning easy and fast by memozing good parameter
settings for past workloads, and a \emph{hierarchical search engine} to
effectively search the parameter space.  Evaluation of \xxx on four main
Linux subsystems shows that it is easy to use and effectively improves
each subsystem's performance.

\end{abstract}

\section{Introduction}

A classic principle in OS kernel design is to separate mechanisms and
policies~\cite{principle}.  Specifically, kernel developers build a small
yet expressive set of mechanisms, on top of which users can implement
flexible policies optimal for their workloads without the need to
re-implement the mechanisms.  In today's monolithic kernels such as Linux,
this principle manifests in a different form.  Specifically, these kernels
often implement a small, fixed set of policies, while exposing many
parameters for users to a policy or adjust the specific settings of a
policy.  For example, Linux provides three I/O scheduling policies named
deadline, cfq, and noop (see \S\ref{diskio} for details), and users can
write a policy's name to \v{/sys/block/<disk name>/queue/scheduler} to
select the policy. These policies each have between 0 to 12 parameters and
users can write to \v{/sys/block/<disk name>/queue/iosched/<parameter
  name>} to change the values of the parameters.


Ideally, the parameters exposed by the kernels should be flexible enough
so that users can tune the parameters to get good performance.  Unfortunately,
this ideal breaks down in practice because of a knowledge gap between
kernel developers and users.  The developers implement the policies so
they know the effects of the parameters well, but they do not know what
workloads users will run.  They typically use workloads that matter to
themselves (\eg, \v{make -j} of the Linux kernel~\cite{kernbench}) to set
the default policies and parameters.  Users know their workloads well but
often lack deep understanding of the kernel internals.  For example, the
Linux completely fair scheduler (CFS) has over 10 parameters with obscure
names tightly tied to the CFS algorithm and implementation.
A ``brief'' Linux performance tuning guide alone has several hundred
pages~\cite{guide}, most of which are just rules of thumb.  Even
performance-tuning experts consider the tuning of OS performance as a
\textit{black art}~\cite{solaris,diagram,tunex}.  Users thus rarely tune
the parameters or tune them correctly, and get stuck with the bad, default
policies and parameters set by kernel developers.  Our experiments show
that these default settings sometimes degrade performance by over 10 times
(\S\ref{diskio}).

\begin{figure}[t]
\centering
\includegraphics[width=0.48\textwidth]{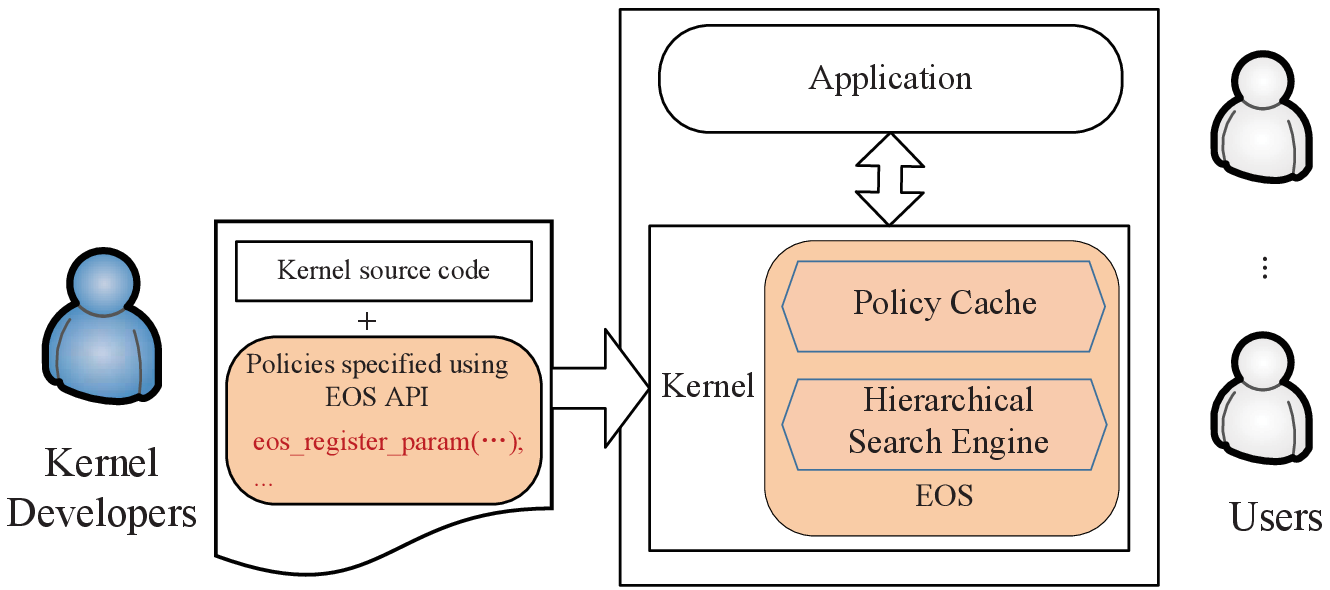}
\caption{\xxx's architecture.} \label{fig:overview}
\end{figure}

This paper presents \xxx, a system that bridges the knowledge gap between
kernel developers and users by automatically evolving the policies and
parameters \emph{in vivo} on users' real, production workloads.  Key in
\xxx are three new ideas, illustrated in Figure~\ref{fig:overview} and
described below.

First, \xxx provides a simple \emph{policy specification API} for kernel
developers to programmatically describe how the policies and parameters
should be tuned while implementing the policies. Specifically, they can
use the API to describe: (1) \emph{metadata} of the parameters that
annotate the parameters with additional information, such as where the
parameters are stored in memory (so that \xxx can modify them) and the
value ranges of the parameters; (2) \emph{sensors} that capture the
characteristics of the workloads, such as the average I/O request size,
which \xxx uses to identify workloads (see next paragraph); and (3)
optimization \emph{targets} that developers intent to measure the system's
performance, such as the number of I/O requests per unit of time, so \xxx
knows what to optimize for. This API helps developers pass their domain
knowledge to \xxx so that it can automatically tune the policies and
parameters in vivo.

Second, \xxx provides a \emph{policy cache} to make in-vivo tuning easy
and fast.  It is often very time-consuming to search a large policy and
parameter space to find a good setting.  Fortunately, workloads are often
stable or repetitive over time.  For instance, Wikipedia HTTP traces show
highly repetitive patterns everyday~\cite{wikibench}; a web proxy I/O
trace at MSR cambridge shows highly stable behaviors~\cite{msr}.  Thus,
once \xxx finds a good policy and parameter setting for a workload, it
stores this setting into the policy cache. Next time a similar workload
comes, \xxx simply reuses the cached setting.  A second use of the policy
cache is to store the intermediate result before the search of a good
setting for a workload is done.  Since there are many settings to search,
\xxx may find a good setting only after the workload has repeated many
times. \xxx stores the intermediate result into the policy cache so that
next time it does not have to restart the search from the very beginning.

Third, \xxx provides a \emph{hierarchical search engine} to effectively
search the policy and parameter space. This framework works as follows.
At the top level, it uses a simple, threshold-based algorithm to detect
which kernel component (\eg, I/O scheduling or page replacement) is the
bottleneck.  Then, at the component level, it enumerates through the
policies and, for each policy, it searches through different values of the
parameters. For each policy and parameter setting, it measures the
system's performance, and picks the best performing setting. \xxx
currently uses a greedy descendant algorithm~\cite{atlas} by default at this
level, and allows developers to plug in their favorite search algorithms.

We explicitly designed our \xxx system to bridge the knowledge gap between
kernel developers and users; it is orthogonal to the massive bodies of
work on creating search algorithms that find good or optimal settings out
of a huge parameter search space. \xxx aims to find a good parameter
setting that significantly improves performance.  The setting does not
have to be the optimal because finding an optimal setting often requires
sophisticated tuning algorithms and is extremely time-consuming.  We
explicitly designed \xxx to handle stable or repeatable workloads, not
flash workloads because flash workloads occur rarely.

We implemented \xxx in Linux.  The policy specification API consists a set
of C macros and functions.  To enable this API, developers simply include
a header file.  The policy cache and the hierarchical search framework are
implemented as a Linux kernel module, dynamically loadable for flexibility
of use.  Its current search algorithms are tailored for local policies
that do not tightly depend on external environments such as networks
because of the unpredictability of these environments.  For instance
policies in the network stack are out of the scope of \xxx.

We evaluated \xxx on the policies in all four main non-networking
subsystems in Linux 3.8.8: disk I/O scheduling, CPU scheduling,
synchronization, and page replacement.  We augmented synchronization and
page replacement with state-of-the-art policies to provide a more thorough
evaluation of \xxx.  Results show that (1) \xxx is easy to use, requiring
a couple of hours and 16 to 50 lines of code to specify the policies in
each subsystem and (2) it effectively improves each subsystem's
performance by an average of 1.24 to 2.58 times and sometimes over 13
times.

The rest of this paper is organized as follows. \S\ref{apc} presents
\xxx's design goals. \S\ref{spec} describes \xxx's policy specification
API, \S\ref{sec:cache} \xxx's policy cache, and \S\ref{sec:engine} \xxx's
hierarchical search engine. \S\ref{sec:imp} discusses implementation
issues. \S\ref{eval} evaluates \xxx's performance. \S\ref{related}
discusses related work, and \S\ref{sec:conclusion} concludes.


\section{Design Goals}\label{apc}

A key design goal in \xxx is to find a parameter settings with good
performance; it does not aim to find the absolute optimal setting.
Massive bodies of work have been devoted to find optimal settings to
maximize system performance. However, current methods for finding optimal
settings are still too complex and too slow. We believe this design goal
makes \xxx ready for practical use by kernel developers and users. In
addition, once advances are made to the algorithms for finding optimal
settings, \xxx can simply adopt them.

\begin{figure*}
\centering
\subfigure[{One-week Wikipedia HTTP trace}]
{
    \includegraphics[width=0.45\textwidth]{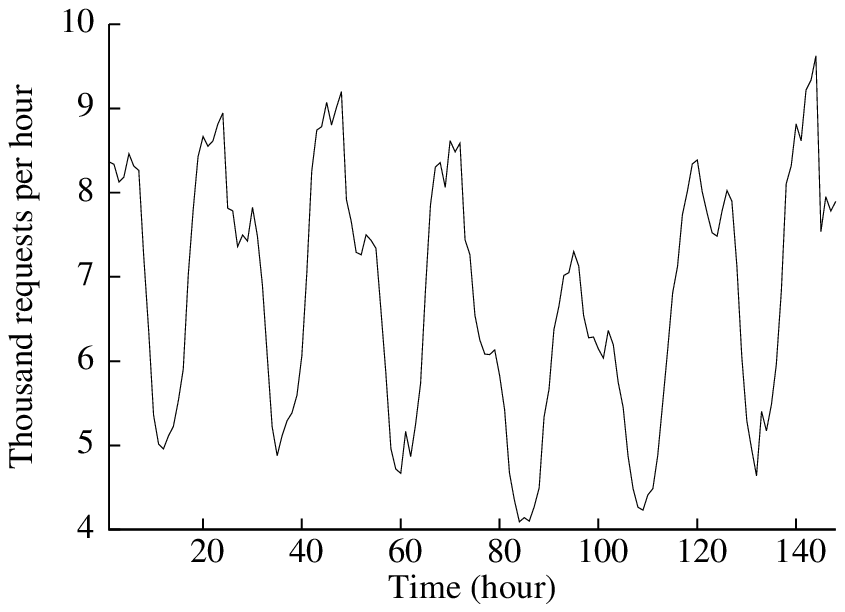}
}
\subfigure[{One-week MSR Web proxy I/O trace}]
{
    \includegraphics[width=0.45\textwidth]{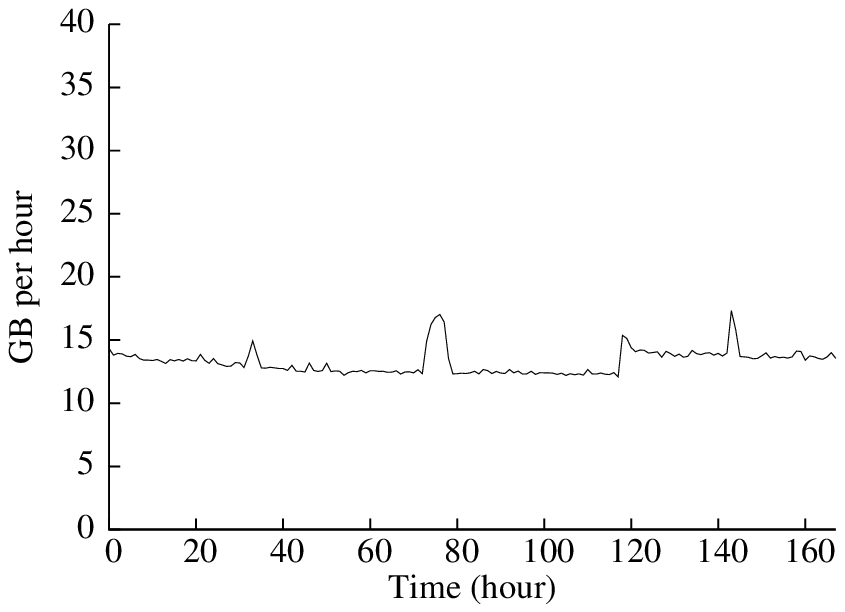}
}
\caption{Traces of two example real-world workloads. The Wikipedia trace
  is periodical, and the MSR trace is stable.}
\label{traces}
\end{figure*}

A second key design goal in \xxx is to focus on relatively steady, repeatable
workloads, not flash workloads.  This design goal benefits \xxx in two
ways.  First, steady, repeatable workloads give \xxx enough time to search
through many policy and parameter settings to find a good setting.  This
search process may take some time, making it difficult to catch up flash
workloads that last for very short periods of time.  Second, steady,
repeatable workloads increase the hit ratio of \xxx's policy cache,
improving its speed.  Recall that the policy cache maps workloads to
settings that yield good performance on the workloads.  The more steady or
repeatable a workload is, the more likely \xxx finds a setting in this
cache for the workload.

In practice, many workloads match this design goal.  Figure~\ref{traces}
shows traces of two such workloads, both collected from real-world server
environments.  The first is a HTTP trace to Wikipedia~\cite{wikipedia},
one of the world's most popular websites. It covers a one-week period,
randomly picked from all traces in WikiBench~\cite{wikibench}, a realistic
web hosting benchmark.  The trace is very periodical, repeating a similar
spiky pattern every 24 hours.  The second trace is an I/O throughput
(measured in GB/hour) trace to a Web proxy server at MSR
Cambridge~\cite{msr}, obtained from the widely used IOTTA trace
repository~\cite{iotta}. The trace is rather steady with few spikes.  It
is unsurprising that many workloads, especially those in server
environments, are steady or repeatable because events that cause flash
workloads are rare.

\section{Policy Specification API} \label{spec}




%


\begin{figure}[t]
\centering
\input codes/api.cpp
\caption{\xxx's policy specification API.}
\label{fig:api}
\end{figure}

Figure~\ref{fig:api} shows \xxx's policy specification API.  A key data
structure in the API is struct \v{eos\_param} which describes the metadata
of a parameter.  It includes the unique name of the parameter, the kernel
subsystem this parameter belongs to (for hierarchical search), the value
range of the parameter (for simplicity, \xxx requires that parameters have
only unsigned long values), how to adjust the value of the parameter in
search (linearly by adding or subtracting 1 or exponentially by doubling
or halving).  In addition, this struct provides a setter and a getter
method for accessing the value of the parameter, and field \v{param}
points to where the parameter is in memory.  As a shortcut, developers can
leave either or both of the methods NULL, and \xxx simply accesses the
\v{param} as a pointer to an unsigned long.
Once a kernel developer fill this struct for a parameter,
she can register this parameter with \xxx with method
\v{eos\_register\_param}.

\begin{figure}
\centering
\input codes/example.cpp
\caption{Annotating a disk parameter in \xxx.} \label{fig:example}
\end{figure}








Figure~\ref{fig:example} shows an example annotating a parameter in the
disk I/O scheduling subsystem. \v{blk\_probe} is a Linux function for
detecting and registering disks.  After a disk is registered via
\v{add\_disk(disk)}, developers use \xxx's API to describe the parameters
that control the I/O scheduling policy for this disk.  Since there are
three disk I/O scheduling policies, the minimum value and the maximum
value of this parameter are 0 and 2 respectively.

\xxx provides a second struct \v{eos\_guard} to capture the dependencies
between parameters.  A parameter may be active only when another
parameter has a certain value.  For instance, the quantum parameter in 
disk I/O subsystem is only active when the scheduler parameter is set 
to CFQ.  Thus, it is useless to tune the quantum parameter unless scheduler 
is set to CFQ.  To
express parameter dependency, a kernel developer passes an \v{eos\_guard}
when registering a parameter.  While in theory a parameter may depend on
multiple other parameters with constraints other than equalities, in
practice we never observed such cases for all parameters in all four Linux
kernel subsystem we evaluated. Thus, \xxx's API supports specifying
dependency only on one parameter with an equality constraint, though
augmenting it is easy.

\xxx requires a third struct \v{eos\_subsys} for supporting a kernel
subsystem. This struct includes three important methods.  First, method
\v{get\_sensors} collects the characteristics of a workload, which the
policy cache uses to distinguish different workloads. The argument
\v{size} specifies the size of \v{sensors} and \v{similarity\_threshold},
two arrays to hold outputs from this method. \v{sensors} holds the
concrete values of the workload characteristics, such as the average size
of all I/O requests. \v{similarity\_threshold} specifies a percentage
variance such that if two sensor values are within the similarity
threshold, they are considered the same.  (See \S\ref{sec:cache} for more
details.)  Second, method
\v{get\_optimization\_target} returns an unsigned long value representing
the subsystem performance on the workload, which \xxx seeks to maximize
when it searches through different parameter settings.  Kernel developers
can choose to let users provide workload-specific optimization targets
by providing a system call for setting the optimization target. Lastly, method
\v{is\_bottleneck} checks whether the subsystem is the bottleneck right
now.  A typical implementation of this method is to compare the percentage
of time the subsystem is busy with a threshold (\eg, 80\%).

Besides the data structures and methods described so far, \xxx's policy
specification API also provides syntactic sugar to further ease the use of
the API.  The most useful syntactic sugar is the macro
\v{eos\_register\_param\_static}: it lets kernel developers register a
\v{eos\_param} struct for a file- or global-scope parameter at compile
time, as opposed to calling \v{eos\_register\_param} at runtime, so that
developers can put this macro right next to the declaration of a
parameter.  To implement this macro, \xxx puts compile-time registered
\v{eos\_param} structs in a special ELF section in the Linux kernel or in
a kernel module using linker script tricks.

\section{Policy Cache} \label{sec:cache}

The policy cache brings two benefits. First, it memoizes good parameter
settings and reuses them on similar workloads, greatly reducing the time
spent in searching for good parameter settings.  Second, it helps make the
search incremental.  Specifically, when a workload runs for a time shorter
than what it takes for \xxx to find a good parameter setting, \xxx stores
the intermediate search result in the policy cache, so that the next time
a similar workload runs, \xxx can resume the search.

\xxx maintains a sub-cache for each kernel subsystem because a subsystem's
parameters are typically independent of another subsystem's.  A sub-cache
is organized as a list of $<$\texttt{workload signature},
\texttt{parameter setting}$>$ pairs, where the workload signature captures
the characteristics of a workload and the parameter setting makes the
subsystem perform well on the workload. \xxx computes the workload
signature by invoking the subsystem's \v{get\_sensors} method.

To search a sub-cache to see if a workload exists, \xxx invokes
\v{get\_sensors} to compute the signature of the workload, denoted $s_1$.
It then scans its list and compares the signature with each signature
$s_2$ on the list to see if $s_1$ is within the similarity threshold of
$s_2$.  For instance, if $s_1$ has a field with value 8, $s_2$'s
corresponding field has a value 10 and similarity threshold 20, then \xxx
considers that these two fields have similar values because 8 is within
20\% of 10.  If \xxx determines that all fields of the two signatures are
similar, it considers the two workloads similar and reuses the setting
associated with workload $s_2$ for workload $s_1$. (There may be multiple
entries matching a given signature; \xxx always returns the first match.)

To store intermediate search results, \xxx stores additional information
used by the search engine (\S\ref{sec:engine}) in addition to the current
parameter setting. When a similar workload runs, \xxx uses the additional
information to resume the search.

Implementation-wise, \xxx limits the cache size to be 1000 and replaces
entries that are least recently used.  In our evaluation, cache
replacement never occurred for any of the workloads.  A 12-hour Wikipedia
trace used only 130 entries, the maximum of all evaluated workloads.  \xxx
persists the policy cache across reboots to \v{/var/eos/cache} to save
warm-up time.

\section{Hierarchical Search Engine} \label{sec:engine}

\xxx's hierarchical search engine operates at two levels: it first detects
which subsystem is bottlenecked and then searches for a good parameter
setting within the subsystem.  Once the subsystem is no longer a
bottleneck, a second subsystem may become bottlenecked and \xxx moves on
to tune the second subsystem.  The rational to focus on one subsystem at a
time matches the general performance tuning experience: performance
problems typically occur when only one subsystem is bottlenecked.  To
detect which subsystem is bottlenecked, \xxx invokes the subsystem's
\v{is\_bottleneck} method.

To search for a good parameter setting, \xxx can in principle leverage the
algorithms proposed by the massive bodies of prior work on performance
tuning~\cite{loukides1990system}. We opted for an algorithm called orthogonal search~\cite{puri1987efficient}
because this algorithm is simple, finishes quickly, and finds parameter
settings with good performance.  Operationally, this algorithm iterates
through the list of parameters and finds the best value for each
parameter.  It then combines these best values into the resultant
parameter setting.xs The intuitions are that (1) parameters are largely
independent so they can be searched separately and (2) the effects of the
parameters on performance are largely monotonic, \eg, if increasing the
value of a parameter improves performance, then we should keep increasing
the value.

We made two modifications to this algorithm to make it more efficient
within the context of \xxx.  First, our modified algorithm prioritizes
toward the more limited parameters -- those with smaller number of
possible values.  The intuition is that, since the number of possible
values is small, the difference in the values often has a large impact on
performance.  Thus, once \xxx finds a good value for a limited parameter
at the beginning of the search process, it enjoys a good application
performance for the rest of the search, improving the average application
performance.  Second, our modified algorithm respects the dependencies
between parameters.  Recall that kernel developers can express when a
parameter is active depending on another parameter using \xxx's API
(\S\ref{spec}). Our algorithm does not search a parameter if it is not
active.

\xxx periodically checks whether it should initiate a new search
process. In our current implementation, this period is every 15 minutes.
When searching within a subsystem, \xxx checks the subsystem's performance
by calling \v{get\_optimization\_target} 5 seconds after it activates a
setting to allow the setting to stabilize.

\section{Implementation}
\label{sec:imp}

We implemented \xxx in Linux 3.8.8.  The policy specification API consists
of a set of C macros and functions.  To enable this API, kernel developers
simply include a header file.  The policy cache and the hierarchical
search engine are implemented as a dynamically loadable kernel module,
making it flexible for users.

To provide a more thorough evaluation of \xxx, we further modified Linux
to add several additional policies to two subsystems.  We describe these
modifications in the next two subsections.


\subsection{Synchronization Subsystem}
 
    
  

\begin{figure}[t]
\centering
\input codes/spinlock.cpp
\caption{Pseudocode for backoff-based ticket lock.}
\label{backoff}
\end{figure}

Linux uses ticket spin lock as the basic low-level synchronization policy.
However, its performance is not good when a lock is seriously contended or
seldom contended~\cite{locks:linuxsymp}. To solve this problem, we made
two modifications to the spin lock implementation, described below.

The first modification adds a back-off to the ticket spin lock.  The
pseudo code is shown in Figure~\ref{backoff}. Instead polling the
\texttt{current} variable in the lock constantly which causes cache line
bounces in high contention, a new lock requester pauses for $C \times N$
before each cache polling, where $C$ represents the polling weight for
each requester and $N$ is the number of the current lock requesters. The
default value of $C$ is set to zero.

The second modification adds two new locking policies to handle low-level
and high-level contention.  These two policies are: (1)
Test-and-Test-And-Set (TTAS) lock, which implements double-checked
locking~\cite{rudolph1984dynamic}; and (2) MCS lock, which lets each lock requester spins on
its own cache line~\cite{mellor1991synchronization}.

We call this new lock implementation the \emph{mixed lock}. Based on a
global variable in the kernel (\texttt{method\_tuner}), the mixed lock
dynamically switches to a lock policy to cope with different contention
levels. This design is motivated by reactive lock~\cite{reactive}.

The data structure of each mixed lock contains four components, including
the data structures of three locks and a \v{mode} field, indicating which
lock policy this mixed lock currently is using.  Note that we cannot use
\v{method\_tuner} to figure out the current lock policy of a mixed lock
because the lock may be already held under a different policy.  All the
lock-policy-specific data structures are stored in the same cache line for
high performance.
The \v{mode} variable is mostly read but seldom written, so it is stored
in a different cache line to reduce false sharing.

The mixed lock interface is as follows.  The functions \texttt{enum
  release\_mode acquire\_mixed\_lock (struct lock* lock, struct qnode
  *node)}, \texttt{enum release\_mode acquire\_mixed\_trylock (struct
  lock* lock, struct qnode *node)}, \texttt{void
  release\_mixed\_lock(struct lock* lock, struct qnode* node, enum
  release\_mode mode)} and \texttt{void init\_lock(struct lock *lock)} are
used to acquire, try to acquire, release and initialize a mixed lock,
respectively.  The function \texttt{acquire\_mixed\_lock(...)} is used to
route the lock request to a specific lock protocol (TTAS, MCS or back-off
based ticket) based on the \texttt{mode} variable of the mixed lock. For
each lock protocol, if the lock is acquired successfully, the
\texttt{method\_tuner} variable is checked to determine whether to switch
to another protocol, or else, the \texttt{mode} variable of the mixed lock
is examined to select the correct protocol to wait. The return value of
\texttt{acquire\_mixed\_lock} indicates whether to switch to a different
lock protocol and acts as a parameter of the lock releasing function.

When releasing a mixed lock, if the \texttt{mode} variable indicates that
we do not need to change lock protocol, \xxx simply releases the trivial
lock by calling the releasing function of the current lock
protocol. 
protocol should be switched to another protocol, \xxx acquires the target
protocol, modifies the \texttt{mode} variable, invalidates current
protocol and finally releases the target protocol.  In this way, we can
ensure that only one valid lock protocol exists and requesters at invalid
protocols will fail the request and retry the valid protocol.

For the three lock protocols in the mixed lock, we give priority to the
TTAS lock by setting it as the default protocol, because most locks in the
kernel are seldom heavily contended. Thus, in function
\texttt{init\_lock}, the \texttt{mode} variable is TTAS and other locks
are invalid. The mixed lock needs a total 300 lines of C code to
implement.

\subsection{Page Replacement Subsystem}
Linux uses LRU2Q as the page replacement policy to determine which pages
should be reclaimed if the free pages are not enough for future use.
Specifically, it maintains two LRU lists, one with pages accessed only
once, and the other with pages accesses more than once.  When a page needs
to be evicted, LRU2Q considers pages on the accessed-once LRU list first.

We implemented another two page replacement policies: (1)
Adaptive Replacement Cache ({\tt ARC})~\cite{megiddo2004outperforming} and
(2) Clock with Adaptive Replacement and Temporal filtering ({\tt
CART})~\cite{bansal2004car}.

Besides the two LRU lists used in LRU2Q, ARC or CART uses two more lists 
(called {\em nonresident lists}) to memorize the pages that are reclaimed from the memory recently. 
When a new page is read into the memory, ARC or CART check whether the page is 
memorized in the nonresident lists and puts the page into appropriate LRU lists. 
For instance, if the nonresident lists show that the page was on the
accessed-more-than-once LRU list, ARC or CART adds the page directly to
this LRU list, bypassing the accessed-once list.

According to previous researches~\cite{bansal2004car,megiddo2004outperforming}, 
if we do not consider the overhead of maintaining the nonresident lists, 
both ARC and CART always perform better than the traditional LRU page replacement policy theoretically. 
However, in practice, because the extra overhead of maintaining the nonresident lists 
is not negligible, ARC and CART can perform worse than LRU2Q (see \S\ref{sec:page} for
details). 

In order to support dynamic policy switch, we isolated the implementation of page replacement policies
with the page replacement mechanism. 
Specifically, we modified and added a set of interfaces into the Linux kernel so that 
\xxx can switch the page replacement policies at runtime. 
Figure~\ref{fig:mimp} gives the functions we implemented in ARC page replacement policy.
Kernel developers can easily develop their own page replacement policies 
following the same pattern in Figure~\ref{fig:mimp}.

\begin{figure}[t]
\centering
\input codes/memory_imp.cpp
\caption{Functions implemented in ARC page replacement policy.}
\label{fig:mimp}
\end{figure}

To switch the page replacement policy, we implement {\tt void pr\_change(struct zone *zone, enum pr\_policy\_list old\_p, enum pr\_policy\_list new\_p)}
that moves all the pages in the LRU lists of {\tt old\_p} to the LRU lists of {\tt new\_p}.
After that, Linux kernel will use the new policy to guard the page replacement in memory.

In \xxx, we set CART instead of LRU2Q as the default policy because we use
the nonresident list hit ratio, which can only be collected in ARC and
CART, as the workload sensor (see \S\ref{sec:page} for details).  One
limitation of this implementation is that once we switch the page
replacement policy to LRU2Q, we cannot switch to other policies.

\section{Evaluation} \label{eval}

We focus our evaluation on the following three questions:

\begin{enumerate}
\item Is \xxx easy to use?
\item Can \xxx find parameter settings that significantly outperform than
  default?
\item Can \xxx consistently benefit different kernel subsystems?
\end{enumerate}

\begin{table}
\center
\begin{tabular}{ccc}
  \textbf{Subsystem} & \textbf{Spec LOC} & \textbf{Average Speedup} \\
  \hline  
  Disk IO scheduling & 50 & 2.58 $\times$ \\
  CPU scheduling & 36 & 3.20 $\times$ \\
  Synchronization & 30 & 1.81 $\times$ \\
  Page replacement & 16 & 1.24 $\times$ \\
\end{tabular}
\caption{\emph{Summary of Results}. The policies in each subsystem require 16 --
  50 lines of code to specify.  \xxx's performance improvements range from
  1.24 to 2.58 times.} \label{tab:summary}
\end{table}

Each of the next four subsections presents a case study of applying \xxx
to one of the four main non-networking subsystems in the Linux kernel:
disk I/O scheduling, CPU scheduling, synchronization, and page replacement
subsystems.  Before we diving into the detailed results for each
subsystem, Table~\ref{tab:summary} summarizes \xxx's performance.  For
every kernel subsystem evaluated, \xxx improved the subsystem's
performance by 1.24$\times$ to 3.20 $\times$ on average.  In addition, to
specify the policies, we only needed to write 16 to 50 lines of code
within a few hours; developers of these subsystems can most likely spend
less time and annotate the parameters better.  These results show that
\xxx is easy to use and can effectively improve performance.

\subsection{I/O Scheduling}
\label{diskio}

\textbf{Specifying policies.} In the Linux I/O scheduling system, we
annotated four parameters, previously reported to have large performance
impact~\cite{ibm,disktest}.  The first parameter specifies which
I/O scheduling policy to use for a disk.  The version of Linux we used
supports three policies: (1) completely fair queuing
(``\v{cfq}'')~\cite{linux-io-cfq}, (2) deadline-based
scheduling(``\v{deadline}'')~\cite{linux-io-deadline}, and (3) first-come
first-served(``\v{noop}'')~\cite{linux-io-noop}.  To change this parameter
(method \v{setter} in struct \v{param}), we used Linux's function
\v{elevator\_change}.  The second parameter specifies the maximum number
of requests on a disk queue.  The third parameter specifies how many pages
to read ahead.  The last parameter specifies the time slice allocated to
each process when the I/O scheduling policy is set to \v{cfq}.  To
identify workloads, we used five sensors: the number of the concurrent
processes, the read-write ratio of the requests, the average I/O request
size, the average seek distance between two consecutive disk requests, and
the average time between two consecutive requests.  These sensors capture
typical workload characteristics that significantly affect performance.
We set the optimization target to be the number of sectors read or written
per second.  In total, it took us roughly 2 hours and 50 lines of code
to specify the policies in the I/O subsystem.

\textbf{Experimental setup.} We used an evaluation machine with Ubuntu
13.04 server edition, a quad-core 3.6 GHZ Xeon E5-1620 processor, 32 GB
memory, and two 2 TB hard disks.  We used seven popular I/O benchmarks and
one 12-hour Wikipedia~\cite{wikipedia} I/O trace as the evaluation workloads.
Specifically, the \v{skip} benchmark~\cite{skip} is a micro-benchmark that
starts multiple processes, each of which repeatedly reads 4 KB from a 2 GB
file and skips the following 4KB.  The \v{vskip} benchmark executes
\v{skip} in a KVM virtual machine to evaluate \xxx's performance in a
virtual environment.  The \v{zmIO}~\cite{zmio} benchmark is another
micro-benchmark that starts multiple processes, each of which sequentially
reads 64 KB from a raw disk using direct I/O.  The TPC-B
benchmark~\cite{tpcb} measures the number of executed transactions per
second for a database.  We used PostgreSQL~\cite{postgresql} as the database and
runs TPC-B on another machine.  The table under test was much larger than
the total memory size on our evaluation machine.  The TPC-H
benchmark~\cite{tpch} is a decision support benchmark; we ran four processes
issuing Q2 requests to a PostgreSQL database simultaneously.  The TPC-C
benchmark~\cite{tpcc} evaluates OLTP applications by simulating an environment
where multiple users submit transactions simultaneously into a database.
The vTPC-C executes TPC-C in a KVM virtual machine.  Parallel grep~\cite{taco}
is a macro-benchmarks that uses the GNU \texttt{grep} utility to look for
a non-existent string in the Linux kernel 3.8.8.  Eight processes are
started simultaneously, but each process calls \texttt{grep} in its own
Linux source code copy to reduce the possible kernel lock contentions.
(We only show results of running \v{skip} and TPC-C in a virtual
environment because the other workloads had no difference executing in a
virtual environment or not.)


 \begin{figure}[t]
\centering
\includegraphics[width=0.48\textwidth]{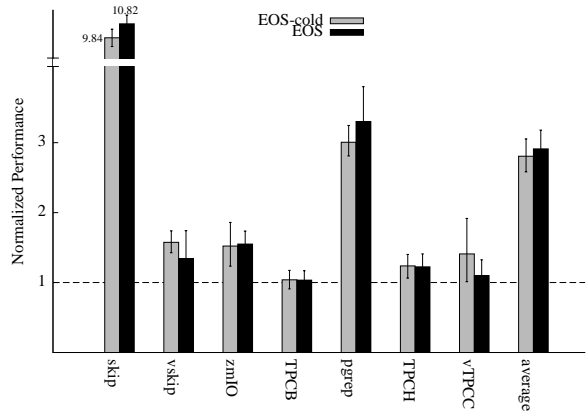}
\caption{\xxx's speedup of disk I/O scheduling subsystem over default
  setting. \xxx-cold represents results with a cold policy cache, and \xxx
  a warm cache. The last two columns show average speedup of all
  workloads.}
\label{disk}
\end{figure}

\textbf{Results.} Figure~\ref{disk} shows the performance improvements of
the benchmarks compared to the default setting (deadline in the version of
Ubuntu we used, three other parameters).  With a cold cache, \xxx achieves
over 11$\times$ speedup on \v{skip} and 3$\times$ speedup on parallel \v{grep}.  
Its improvements on other applications are also quite significant, up to 41\%. 
With a warm cache, \xxx gains another 10.0\% on average. Since we designed 
\xxx to handle relatively steady, repeatable workloads, we expect that the 
warm-cache case is more often than cold-cache.

\begin{figure}[t]
\centering
\includegraphics[width=0.48\textwidth]{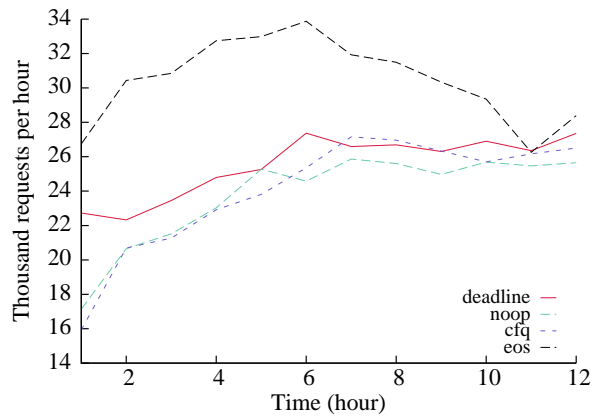}
\caption{\xxx's speedup of disk I/O scheduling subsystem on a 12-hour
  Wikipedia trace.}
\label{tw}
\end{figure}

Figure~\ref{tw} presents \xxx's performance on a 12-hour Wikipedia trace.
It outperforms the default policy by over 19.8\% for almost the entire
workload.

\begin{figure}[t]
\centering
\includegraphics[width=0.48\textwidth]{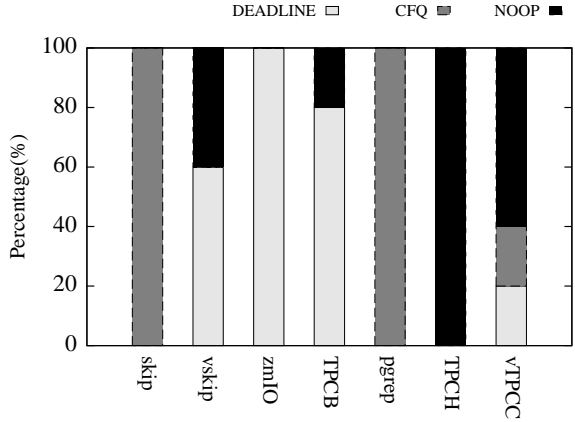}
\caption{Percentage of time a disk I/O scheduling policy is selected by \xxx.}
\label{selected}
\end{figure}

Figure~\ref{selected} shows the parameter settings \xxx selected for
each workload.  Note that \xxx may select more than one settings over
repeated executions when the settings have roughly the same performance.
The figure also shows that there is no single scheduling policy that fits
all workload.


\begin{figure}[t]
\centering
\includegraphics[width=0.48\textwidth]{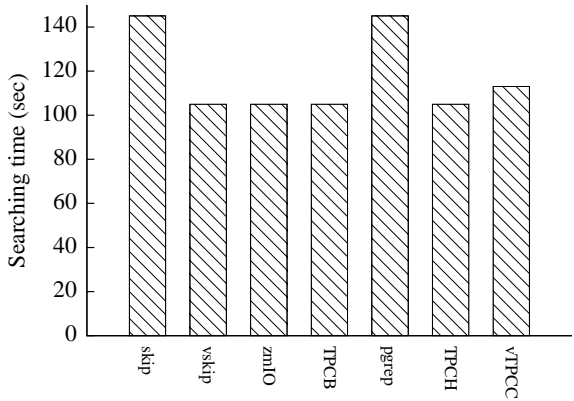}
\caption{Search time for each disk I/O scheduling workload.}
\label{searchtime}
\end{figure}

Figure~\ref{searchtime} shows the time it takes for \xxx to find a good
setting for each workload.  This time ranges from 105 seconds to 145 seconds.

\begin{figure}[t]
\centering
\includegraphics[width=0.48\textwidth]{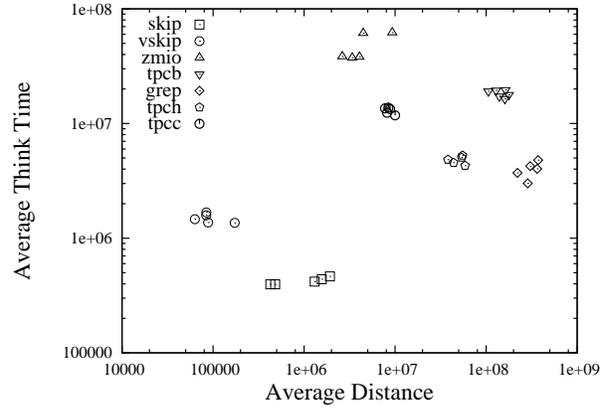}
\caption{Workload clusters based on two sensors: average distance and average
  time between two consecutive I/O requests.}
\label{diska}
\end{figure}

\xxx uses workload signatures to search for settings in the policy cache,
so we also studied the differences and similarities of the signatures of
the evaluated workloads.  Figure~\ref{diska} plots the average seek
distance vs the average time between two consecutive disk requests for the
workloads over repeated executions.  These two sensors suffice to place
each workload in its own singleton cluster.


\subsection{CPU Scheduling}
\textbf{Specifying Policies.}  In the CPU scheduling subsystem, we
annotated three parameters. Specifically, \v{sysctl\_sched\_latency\_ns}
specifies the length of time for scheduling each runnable process once.
\v{sysctl\_sched\_min\_granularity\_ns} specifies the minimum time a
process is guaranteed to run when scheduled.
\v{sysctl\_sched\_wakeup\_granularity\_ns}, it describes the ability of
processes being waken up to preempt the current process. A larger value
makes it more difficult preempt the current process~\cite{cfstunable}.  To
identify workloads, we used one sensor: the number of retired instructions
executed in user space. We set the optimization target to be the same.
Overall, it took us 2 hours and 36 lines of code to specify the policy.
    
\begin{figure}[t]
\centering
\includegraphics[width=0.48\textwidth]{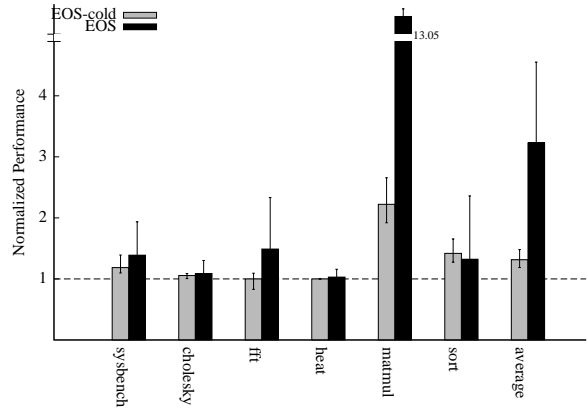}
\caption{\xxx's speedup of CPU scheduling subsystem over default setting.}
\label{scheduling}
\end{figure}

\textbf{Experimental Setup.}  The evaluation machine is the same as the
one used to evaluate the disk I/O subsystem.  To demonstrate the
performance improvements of our system, we used the CPU scheduling
benchmark in SysBench, a widely used systems benchmark~\cite{sysbench} and
five parallel computing benchmarks in the popular Cilk~\cite{cilk}
benchmark suite.  SysBench starts multiple threads, each of which
repeatedly locks a mutex, yields the CPU, and unlocks the mutex.  The Cilk
benchmarks run basic algorithms such as merge sort and fast Fourier
transformation in parallel.

\textbf{Results.}  Figure~\ref{scheduling} shows \xxx's performance
improvements compared to the default.  With a cold cache, \xxx worked well
for most workloads and achieved more than 2$\times$ speedup on matmul.
With a warm cache, it performed even better, achieving a 13$\times$
speedup on matmul.  To better understand this huge speedup, we analyzed
the executions of matmul. It turns out that, during the execution of
matmul, sometimes many threads have no work to do and are busy trying to
steal work from other threads without yielding the CPU.  These futile
work-stealing requests waste many CPU cycles.  In this scenario, \xxx
correctly adjusted the CPU scheduling parameters to preempt threads more
often so that the few threads with work to do can make good progress,
significantly improving performance.

It took \xxx 200 seconds to finish the search for each workload. (There is
only one CPU scheduling policy in the version Linux we used, so \xxx
adjusted the parameters of this policy only.)  We also studied the
signatures of the workload.  Based on the signatures, the workloads fall
into three clusters: (1) SysBench, (2) fft, (3) heat, and (4) the rest of
the workloads.

\subsection{Synchronization}
\label{syn}

\textbf{Specifying policies.}  In the synchronization subsystem, we
annotated two parameters. \texttt{method\_tuner} specifies which locking
policy to use.  \texttt{val\_tuner} specifies the polling weight in the
back-off ticket spin lock ($C$ in Figure~\ref{backoff}).  \v{val\_tuner}
is active only when \v{method\_tuner} is set to be back-off based ticket
lock. To identify workloads, we used one sensor: the average lock
acquisition time (the time between a lock is requested and the lock is
granted). We used the number of lock acquisitions per second as the
optimization target.  In total it tooks us 2 hours and 30 lines of code to
specify the policies in the lock subsystem.



\textbf{Experimental Setup.}  We use an evaluation machine with Ubuntu
13.04 server edition, eight 48-core 1.9GHZ AMD Opteron processors.  We
used four benchmarks as workloads.  Specifically, fops~\cite{boyd2012non} is a
micro-benchmark that measures the locking performance of the Linux
directory entry cache.  It starts multiple processes, each of which
repeatedly opens and closes a private file.  The mmapbench benchmark is a
micro-benchmark written by us to measure the locking performance of the
Linux memory map operation.  It spawns multiple processes, each of which
repeatedly maps the same continuous 500 MB in shared, read-only mode,
reads the first byte, and destroys the mapping.  Parallel
postmark~\cite{taco} is a macro-benchmark that simulates file server
workloads hosting email and news services.  It starts multiple threads,
each of which runs file system operations repeatedly on an independent set
of files (between 0.5 and 10K bytes in size).  To measure locking instead
I/O performance, we used the \v{tmpfs}.  dbench~\cite{dbench} is a popular
macro-benchmark for benchmarking file systems.  It starts multiple
processes, each of which does many file operations such as read, write,
link, and unlink.  We also used \v{tmpfs} for dbench.

\begin{figure}[t]
\centering
\includegraphics[width=0.48\textwidth]{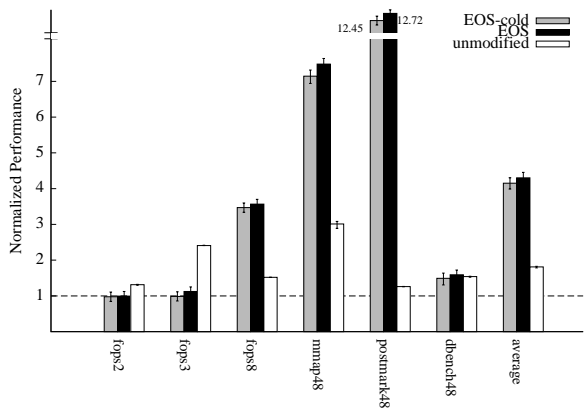}
\caption{\xxx's speedup of synchronization subsystem over default
  setting.\xxx-cold represents results with a cold policy cache, \xxx a
  warm cache, and unmodified the results of the unmodified Linux spin lock
  implementation. }
\label{lock}
\end{figure}

\textbf{Results.}  Figure~\ref{lock} presents \xxx's performance
improvements over the default.  To create different lock contention
levels, we varied the number of processes or threads used by each
benchmark. For example, fops8 means running fops with eight processes.
With a cold cache, fps8, mmapbench, and parallel postmark performed much
better with \xxx than the default (up to 13$\times$).  With a warm cache,
\xxx performed even better.  (Since we modified Linux's spin lock to
support multiple locking policies, our code adds some overhead.  Thus,
Figure~\ref{lock} shows the performance of Linux's unmodified spin lock
for reference.)

\begin{figure}[t]
\centering
\includegraphics[width=0.48\textwidth]{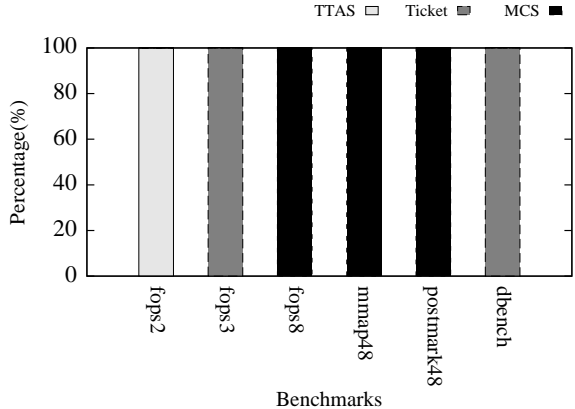}
\caption{Percentage of time a lock policy is selected by \xxx.}
\label{lock_s}
\end{figure}

Figure~\ref{lock_s} shows the locking policy \xxx selected for each
workload.  For every workload, \xxx consistently selected the same policy
for the workload over repeated executions.  When the contention level is
low (fops2), it selected TTAS.  When the contention level is medium (fops3
and dbench), it selected the back-off based ticket lock.  When the
contention level is high (fops8, mmapbench, and parallel postmark), it
selected MCS.  The figure also shows that there is no single locking
policy that fits all workload.

It took \xxx 15 seconds to finish the search for fops2, fops8, mmapbench,
and parallel postmark; and 75 seconds for fops3 and dbench. We also
studied the signatures of the workloads.  Based on the signatures, each
workload falls into its own singleton cluster.


   
\subsection{Page replacement}
\label{sec:page}

\textbf{Specifying policies.} 
In the Linux memory management system, we annotated one parameter 
that specifies which page replacement policy to use for memory. 
To identify workloads, we used one sensor: the hit ratio of the
nonresident lists.  This sensor characterizes a performance-critical
memory access pattern of a workload.  Specifically, when a hit occurs on a
page on a nonresident list, the kernel can use access history stored in
the list to effectively improve performance.  We set the optimization
target to be the number of page-in or page-out operations per second.  In
total, it took us roughly 2 hours and 16 lines of code to specify the
policies in the memory management subsystem.

\textbf{Experimental setup.} We used an evaluation machine with Ubuntu
13.10 desktop edition, a quad-core 3.4 GHZ i7-2600 processor, 4 GB memory,
and one 1 TB hard disk.  We used one synthetic micro-benchmark and six
memory-intensive benchmarks for big data area and scientific computing
area.  Specifically, WordCount and TeraSort are example programs in Hadoop
package~\cite{hadoop}. WordCount counts the number of different words
appear in a 10GB file.  TeraSort sorts a large number of numbers (overall
size is 10GB) using merge sort. Join~\cite{bigdatabench} is one of the
popular big data benchmarks developed based on hive~\cite{thusoo2009hive}.
It does join operation on two large tables in a mysql database.
Heat~\cite{LAWS} simulates the heat distribution over time on a metal
plate.  Stencil~\cite{LAWS} does the 9-point iterative stencil computing.
Sor~\cite{LAWS} is the successive over-relaxation algorithm.
\v{synthetic} is a micro-benchmark we developed. it requests 3.8GB memory
space and writes random data to the space round-and-robin for 10 times.
The data set of all the above benchmarks are larger than the memory size
so that the page replacement happens frequently.

\begin{figure}[t]
\centering
\includegraphics[width=0.48\textwidth]{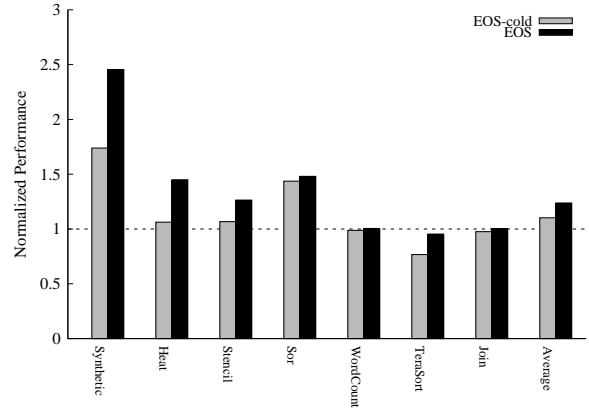}
\caption{\xxx's speedup of the page replacement subsystem over default setting.}
\label{fig:memory}
\end{figure}

\textbf {Results.}
Figure~\ref{fig:memory} shows the performance improvements of the benchmarks compared to the 
default setting.
With a cold cache, \xxx achieves 1.74$\times$ speedup on {\tt synthetic}.
With a warm cache, \xxx gains another 72\% on {\tt synthetic}. 
On real world application {\tt Sor}, \xxx achieves 1.43$\times$ speedup with 
a cold cache and achieves 1.48$\times$ speedup with a warm cache.
On average, \xxx can speed up the applications by 1.24 times.

For every workload, \xxx consistently selected the same policy
for the workload
over repeated executions. It selected CART for Heat, Stencil, Sor, and
\v{synthetic}, and the default LRU2Q for WordCount, Terasort, and
Join. (ARC is not selected for any workload.) The figure also shows that
there is no single page replacement policy that fits all workload.

It took \xxx 15 seconds to finish the search for each workload.  We also
studied the signatures of the workload.  Based on the signatures, the
workloads fall into three clusters: (1) Heat, (2) Stencil and Sor, and (3) WordCount, TeraSort and Join.

\section{Related Work} \label{related}

Massive bodies of work has been devoted to improving performance.
However, to our knowledge, no prior work proposed the idea of a
programming API for kernel developers to describe the policy parameters;
nor did any prior work proposed the idea of a general policy cache for
memoizing good parameter settings and reusing them on similar workloads.
Below we compare \xxx to the closely related work.

\noindent\textbf{Performance auto-tuning.}  Self-adapting operating
systems~\cite{hotos} aim to collect traces from real workloads and run
simulations with the traces to adapt the operating system implementation,
not just the parameters. While this goal is exciting, we are not aware of
any implementation of such a system.

Many performance tuning algorithms have been proposed, some of which may
be incorporated to \xxx's search engine for searching a good parameter
setting within a kernel subsystem.  For example, IBM researchers use
generic algorithms to search for the optimal parameter setting for the
Anticipatory I/O scheduler and the Zaphod CPU
scheduler~\cite{generic}. Using several synthetic benchmarks, they
improved performance by 9\% on average.  \xxx improved performance much
more potentially because it can (1) cache prior tuning results and (2)
tune not only the specific settings of a policy, but also which policy to
use.

Borrowing from control theory, feedback-based tuning algorithms
iteratively adjusts parameter settings based on some form of
feedback. Reactive lock~\cite{reactive} dynamically selects the correct
locking protocol based on the lock contention level; our lock
implementation is motivated by reactive lock and further generalizes it by
adding a back-off based ticket lock for mid-level contention.  Application
heart beat~\cite{heart} is one mechanism for application to provide custom
feedback. It is shown to make Linux's CFS scheduler to deliver predictable
performance and temperature~\cite{dac}.  Berkeley Tessellation kernel uses
feedback control to allocate an optimal number of cores for an
application~\cite{dacber}. Feedback-based algorithms tend to take many
iterations, so \xxx's policy cache can help them avoid costly tuning when
an optimal setting already exists in the cache.


Another approach is model-based performance tuning where researchers
construct a performance model for a real system, trains the model
parameters with offline simulation, and applies the model online to select
optimal parameters.  However, model construct typically requires deep
expertise, and is time consuming to build.  One case study took over a
year to optimize four parameters in Berkeley DB~\cite{diagram}.  \xxx's
policy cache can be viewed as a weak, automatically constructed model that
captures a partial mapping from workloads to good parameter settings.

\noindent \textbf{Ad hoc parameter caching.}  Three prior systems touch
upon the idea of caching good parameter settings for certain workloads,
but in an ad hoc way.  One system caches good intermediate ``genes'' to
speed up genetic algorithms~\cite{finger}.  Linux caches congestion
control window sizes based on IP addresses~\cite{sarolahti2002congestion}.  Another system caches
the optimal number of virtual machines allocated for an online service
based on hardware performance counters~\cite{asplos}.  The caching in
these systems is limited to a particular algorithm or parameter, whereas
\xxx supports generalized policy caching.

\section{Conclusion} \label{sec:conclusion}

We have presented \xxx, a system that bridges the knowledge gap between
kernel developers and users by automatically evolving the policies and
parameters \emph{in vivo} on users' real, production workloads. It
provides a simple \emph{policy specification API} for kernel developers to
programmatically describe how the policies and parameters should be tuned,
a \emph{policy cache} to make in-vivo tuning easy and fast by memozing
good parameter settings for past workloads, and a \emph{hierarchical
  search framework} to effectively search the parameter space.  Evaluation
of \xxx on four main Linux subsystems shows that (1) it is easy to use,
requiring 16 to 50 lines of code to specify the policies in each
subsystem, and (2) it effectively improves each subsystem's performance by
an average of 1.24 to 2.58 times and sometimes over 13 times.




\bibliographystyle{abbrv} 
\bibliography{ref}

\end{document}